# Opening Scholarly Communication in Social Sciences by Connecting Collaborative Authoring to Peer Review

*Afshin Sadeghi, Johannes Wilm, Philipp Mayr, Christoph Lange*

The objective of the OSCOSS research project on "Opening Scholarly Communication in the Social Sciences" is to build a coherent collaboration environment that facilitates scholarly communication workflows of social scientists in the roles of authors, reviewers, editors and readers. This paper presents the implementation of the core of this environment: the integration of the Fidus Writer academic word processor with the Open Journal Systems (OJS) submission and review management system.

## Introduction

The objective of the DFG-funded OSCOSS research project (Mayr and Lange 2016) on "Opening Scholarly Communication in the Social Sciences" is to build a coherent collaboration environment that facilitates scholarly communication workflows (Sompel et al. 2004) of social scientists in the roles of authors, reviewers, editors and readers. A collaborative writing environment (Whitehead 2005), for which we chose the Fidus Writer academic word processor[1], is the hub of the overall environment. Further components that we are currently integrating include databases of metadata about scientific publications and research data sets; we are also planning to include repositories hosting the source code of data analysis software. This paper presents the integration of Fidus Writer with a submission and review management system; for this, we chose Open Journal Systems (OJS)[2].

In the "Background" section, we give an overview about the scholarly authoring and reviewing process, our intended approach and the supported workflows including authoring workflow, journal/conference editing workflow and reviewing workflow. The "Related Work" section describes popular state of the art collaborative writing and reviewing systems. In the "Requirements" section we outline the requirements from the perspective of authors, editors and reviewers. The integration of Fidus Writer and OJS is technically described in "Implementation". The paper ends with our "Conclusion".

## Background

Manuscripts submitted to journals and conferences are typically written using word processors. A breakthrough in this area in recent years has been a move from traditional desktop applications to collaborative, web-based versions. Modern online platforms such as Google Docs[3], ShareLaTeX[4] and Overleaf[5] have made collaborative authoring significantly easier. The entire writing process can be completed online – from the writing of the first draft to the production of a camera-ready document. On the other side of the life cycle of scientific articles is the reviewing workflow (Bornmann 2011; Bornmann and Daniel 2010). Mature submission and review management systems supporting this workflow, including the open source solution Open Journal Systems (Smecher 2008) and the centrally hosted solution EasyChair[6] which follows a free usage model; both are web-based platforms.

---

1  https://www.fiduswriter.org
2  https://pkp.sfu.ca/ojs
3  https://docs.google.com/
4  https://de.sharelatex.com/
5  https://www.overleaf.com

These two processes together form the main parts of scientific writing, and they are intertwined, as, after reviewing, the control over a manuscript is typically given back to the authors, who will then revise it. Nevertheless, to our knowledge there is currently no mature system that supports both steps in combination. Therefore, the reviewing workflow is typically realized in a way that requires manual file transfers at several points: after the authoring process is finished, the authors have to export a copy of the manuscript from the collaborative authoring system and submit it to the submission and review management system. There, the reviewers have to download the manuscript, comment on it, upload it back to the system, which will notify the journal editor or the conference PC chair, who will take a decision and notify the authors, who read the comments and, unless rejected, applies revisions to the manuscript before submitting it once more. This review cycle can be repeated two or more times. In every such cycle, authors as well as reviewers have to download and upload a document, authors have to apply revisions to a document according to comments that are given as plain text or, in the best typical case, as annotations to a copy of the document itself. Unless the manuscript is submitted in an editable format, such as an office word processor format, but in a read-only format such as PDF, the authors have to apply revisions by viewing side-by-side their own document in their word processor and the reviewers' annotations in a document viewer application. This procedure is error-prone for authors, as they may overlook comments or apply revisions in the wrong place of the document. It is also cumbersome for the majority of reviewers who are not using PDF annotation tools but write even minor revision requests into their plain-text overall summary of their review, as they have to refer to texts by approximate references such as "in the 2nd paragraph of page 7".

## Approach

With the OSCOSS platform, we aim at linking the two steps of authoring and reviewing scholarly articles. The omnipresence of the Web means that the involved systems already have ways to connect. We choose the free open source systems Fidus Writer and Open Journal Systems (OJS), a submission and review management system, as they have open plugin APIs. This means that all that is left is extending the functionality of each system by a plugin capable of communicating with the other system, e.g., via RESTful web service interfaces. This not only allows the submission of the articles directly to the reviewing environment – it even lets authors and reviewers interact directly on specific sections of an article, which is much more fine-grained than peer review is typically done at the moment.

---

6   http://easychair.org

## Supported Workflows

In this article, we use the term "workflow" to denote a largely predetermined procedure that users in different roles will take through the framework. It varies slightly from journal to journal and from conference to conference.

In this section we will describe the workflows that we considered necessary to be addressed by our implementation of the OSCOSS platform.

**Authoring workflow**. An author starts writing about his research and invites his collaborating researchers to participate. Each person could be responsible for a part of the manuscript; some authors might just review text written by other authors. The corresponding author submits the article to a conference or journal. After the review, the authors receive feedback, usually including some concrete advice on how to correct or update parts of the writing.

**Journal/conference editing workflow.** The journal editor or conference PC chair assigns one, typically two to three reviewers to review every manuscript submitted. The editor also takes the final decision on whether to accept or reject a manuscript or to request the authors to resubmit a revision. He makes this decision based on the review of the reviews or sometimes his own, additional review of the manuscript.

**Reviewing workflow**. The reviewing process starts when a reviewer opens an assigned manuscript. His role is to assess the quality of the manuscript. He is expected to appraise the manuscript as a whole but can also give fine-grained feedback on specific issues in specific places of the manuscript. The reviewer may provide the authors with constructive feedback suggesting how they could improve, shorten or extend the manuscript. Finally, the reviewing workflow includes writing general feedback to journal editor to help him to quickly decide on whether to accept or reject the manuscript or to request a revision from the authors. Depending on the policy of the conference or journal, this cycle can be repeated several times.

## Related Work

There are several collaborative writing and reviewing systems available. In our inventory we found that the existing systems all have limitations.

Among existing collaborative writing systems are: Microsoft Office Online [7], Zoho Office [8], Etherpad [9], Google Docs [10], ShareLaTeX [11], Overleaf [12], and Authorea [13] (see Table 1).

Fidus Writer, the word processor of our choice, combines the advantages of several of these systems: It has a classic word processor interface enabling also non-technical people to use it; Fidus Writer is developed in Python and JavaScript and is open source. It can be installed locally, it offers scientific features such as citation and figure management in a simple What-You-See-Is-What-You-Get (WYSIWYG) interface, yet the editing interface only allows semantic rather than stylistic changes to the text. This means that users can specify that a certain part of the text is a headline of the second level, a link, or emphasized text, but the user cannot specify the font, font size or line height the way word processors allow it. This removes a number of uncertainties such as the question of whether a bold text by itself on a single line should be interpreted as an emphasized paragraph or a small headline. It also removes the problem of users misusing the interface – for example by manually entering 25 line breaks to obtain a page break. The fact that the text is available in its semantic form at all times means that it can be converted fully automatically without the need of human interpretation of its meaning and it can therefore easily be dealt with in any publishing pipeline -- even directly to PDF in a web browser with JavaScript and CSS (Murakami and Wilm 2015). Receiving the text in an unclean format where the semantics are open to interpretation is a major problem for publishers (Wilm 2015).

---

7 https://products.office.com/en/office-online
8 https://www.zoho.com/docs/office-suite.html
9 http://etherpad.org
10 https://docs.google.com
11 https://www.sharelatex.com
12 https://www.overleaf.com
13 https://www.authorea.com

Other existing collaborative academic authoring systems range from ShareLaTeX and Overleaf (Perkel 2014), which allow for the editing of LaTeX code online and are used mainly by programmers and other technically minded people who do not mind reading and writing in code, to Google Docs, Zoho Office and Microsoft Office 365 (both Desktop and Online). These systems offer an interface fairly similar to the desktop version of the traditional Microsoft Word most users are familiar with, which makes them more usable for non-tech experts, but they lack features specific to scientific writing, and they produce output that is similarly complex to deal with in the publishing process as any desktop word processor.

Among the common limitations of such systems there is the software license and terms of use. Most systems store their documents on the servers of the company that operates them; they are typically not available for local installation, which would allow for a custom configuration and keep sensitive data, e.g., of medical studies, within the scope of the security and data protection policies of the authors' organisations. ShareLaTeX is different in that it is open source and can be installed locally. Etherpad is another example of an open source collaborative text editor, but we did not consider it in our comparison as it lacks scientific features.

OJS with a history of more than 10 years from the first release is the submission and review management system of our choice[14]. It is open source and is implemented in PHP. At the time of developing our integration code, OJS version 3 was under development; therefore, our implementation is based on this version. At the time of writing this article, more than 10,000 journals used OJS[15]. In version 3, OJS features a more dynamic interface than previously and supports different peer review configurations including single blind and double blind reviewing. It has an extensive documentation and free tutorials available.

To the best of our knowledge, none of the authoring and reviewing systems mentioned above have ever been integrated with each other. The only existing solution is the ARPHA Writing Tool[16], which is trying to create both a writing and reviewing system. In comparison to Fidus Writer, ARPHA authoring tool does not support realtime collaborative editing. Their reviewing system which is used by less than a handful of publishers is not as mature as OJS in terms of documenting and ease of use. In the next section we explain the requirements that we considered for an integrated system.

---

14  Other existing submission and review systems are: EJPress, EPRESS, ePublishing Toolkit, ESPERE, FontisWorks, OJS, SmartPublishing, Rapid Review, ScholarOne, Editorial Manager, etc.
15  https://pkp.sfu.ca/ojs/ojs-usage/ojs-stats/
16  http://arphahub.com/

| System / Feature | Open source | Academic content | WYSIWYG | Export formats |
|---|---|---|---|---|
| Google Docs | No | Formulas | Yes | DOCX, ODT, PDF, HTML |
| Microsoft Office Online | No | - | Yes | DOCX, ODT, PDF |
| Zoho Docs | No | Formulas | Yes | DOCX, ODT, PDF, HTML |
| ShareLaTeX | Yes | Formulas, Citations | No | LaTeX, PDF |
| Overleaf | No | Formulas, Citations | No | LaTeX, PDF |
| Authorea | No | Formulas, Citations | Yes | DOCX, LaTeX, PDF |
| Etherpad | Yes | - | Yes | HTML |
| Fidus Writer | Yes | Formulas, Citations | Yes | DOCX, ODT, PDF, LaTeX, HTML |

Table 1: Collaborative online writing systems

## Requirements

From the perspective of authors, editors and reviewers, an integrated system needs to comply with certain requirements to be useful. By reflecting on our own experience in each of these three roles, and by talking to the editors of the GESIS journals mda[17] and HSR[18], which will serve as pilots for evaluating our implementation in the scope of the OSCOSS project, we obtained the following list of requirements.

• Ease of use. The integration of the two systems is supposed to make the two workflows easier, so it should reduce the number of manual steps instead of increasing the complexity of the workflow by adding steps.

• Continuity of the workflows through the two applications. When a function of one application requires to call a function of the other system, this call must be performed in the background. Jumping from one system to the other must appear seamless to the author. A user who is registered in one system must also be known in the other one (single sign-on). When a reviewer logs into OJS, he must also be able to see his assigned manuscripts in Fidus Writer without having to log in a second time.

• Each system must support its part of the peer review process. For example, comments whose visibility is restricted by role (e.g. reviewers' comments who are only visible to the authors once approved by the editor), must be available in Fidus Writer, but none of the currently existing peer review services of OJS should be stopped from working; therefore, OJS still needs to support users who are not using Fidus Writer.

---

17  http://www.gesis.org/angebot/publikationen/zeitschrift-mda/
18  http://www.gesis.org/hsr/aktuelle-hefte/

• Conformance to technical standards e.g. RESTful API best practices guidelines, OAuth standards, etc.

• Support the import of manuscripts in wide used formats such as Microsoft Word. This is crucial to attract authors with a background of using different authoring systems, as it enables them to continue working on manuscripts they have started drafting outside of our integrated environment.

• The system needs to be able to export to a variety of formats, such as Microsoft Word, PDF and LaTeX. This is important to support the publishing workflows of as many publishers as possible.

• Support widely used conference and journal publishing templates and layouts, e.g., those of ACM or Springer.

• Handling of graphics, diagrams and tables. Graphics and Tables are present in almost of articles. Therefore support for formulas and tables is crucial as a usability factor for the integrated system.

• Security. The interactions of the Fidus Writer and OJS side of the system over the network should introduce no new security flaws.

## Implementation

OJS already provides online support for the reviewing workflow. In the classic workflow, authors register in OJS system and deliver their article in the form of a PDF or Word file upload. Reviewers have to download the document, review it and give their general feedback to authors and journal editors. Given that OJS is extensible via a plugin mechanism, we were able to extend it with a RESTful interaction API without changing any core OJS code. Representational state transfer (REST) Web services are a method of providing interaction between computer systems on the Internet using a uniform and predefined set of stateless operations.

Our first change to the conventional review workflow affects authors. They no longer need to register on the OJS site. By their submission of a new article through the Fidus Writer interface, we register them as the corresponding author on the OJS site. If the authors are already known to OJS, we link the new article to the previously registered author in the system. And while reviewing, sending the general feedback of the reviewers, i.e. the one that addresses the manuscript as a whole rather than specific parts, to the registered email addresses of the authors makes no need for authors to visit the OJS altogether.

Fidus Writer already supports a collaborative authoring workflow. It allows multiple users to edit a document at the same time (Wilm and Frebel 2015). It allows authors to discuss a general topic among each other using a chat interface. It also supports comments to tag and do internal review during of the draft in the authoring phase. To support the reviewing workflow, we extended Fidus Writer by adding roles to it. The reviewing workflow adds different roles of users to Fidus Writer, each having specific permissions. During the review process, comments can have different levels of visibility. Based on the policy of our targeted journals, we added the permission mechanism based on roles into Fidus Writer. For example, authors see the comments of authors and reviewera, but each reviewer does not see the comments from other reviewers. It was necessary to match user roles and permission levels of Fidus Writer to those of OJS to make them compatible, as shown in Table 2.

| OJS | FW |
|---|---|
| Author | Author |
| Reviewer | Reviewer |
| Journal Editor | Admin |

Table 2: Mapping of roles between OJS and Fidus Writer (FW)

In both OJS and Fidus Writer, roles are defined per document and can differ across documents. While a role in OJS is customizable, in Fidus Writer, the same user can only have one role per document and the permissions of each role are set. For example, a user can be an author, reviewer and the journal editor at the same time in the OJS but he can only have one role in Fidus Writer in relation to one document. Figure 1 shows the matching of users and the roles and how the definition of a role affects the role of a user in the system at the other end. The role of an author in Fidus Writer causes the creation a new author role in OJS and the creation of a reviewer role in OJS causes a similar role to be assigned in Fidus Writer. This allows Fidus Writer to authenticate reviewers via our extension in OJS. As the number of journal editors and administrators is usually limited for the two systems, we did not introduce admin and journal editor roles within the integrated system.

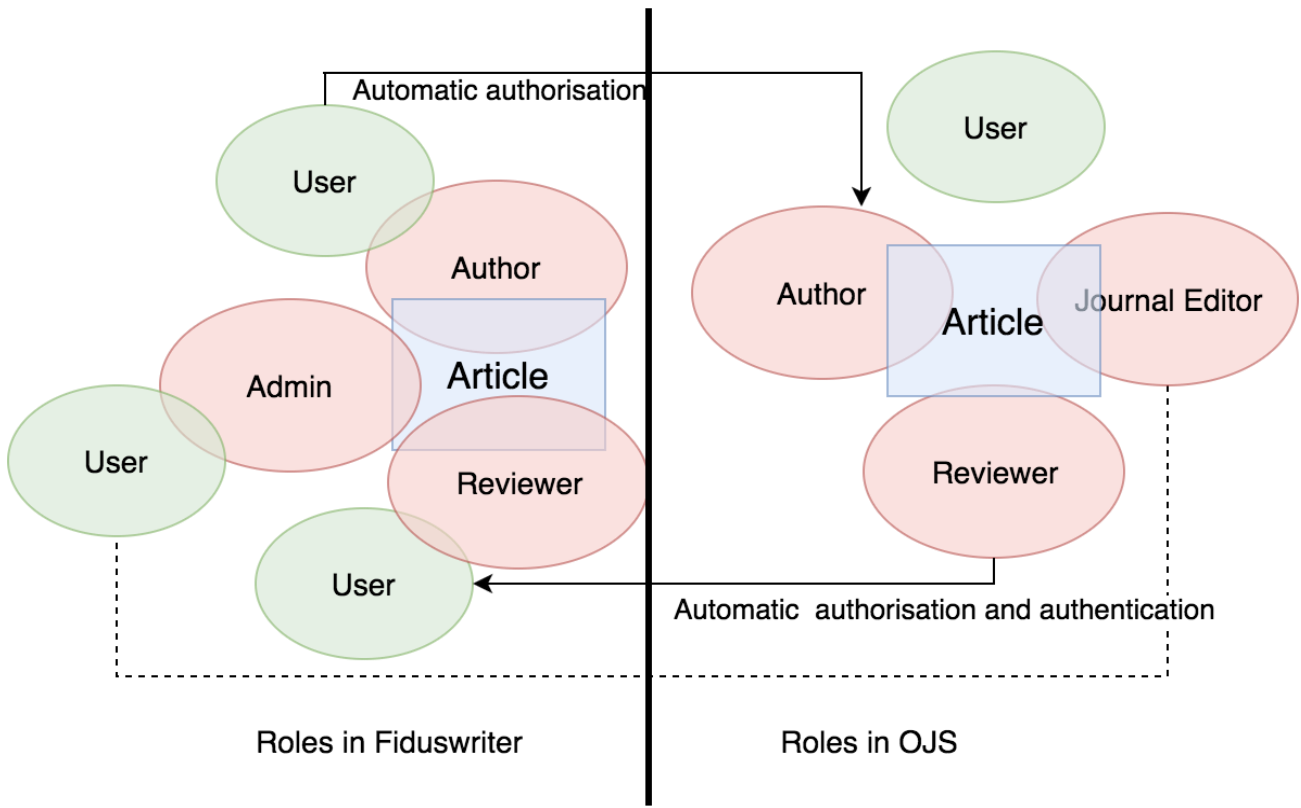

Figure 1: Matching of users and the roles in Fidus Writer and OJS

## Architecture of the integrated system

Fidus Writer is implemented based on Django framework for Python and features a JavaScript rich client and therefore follows the model view controller pattern closely. The controllers are such configured that allow internal interaction based on HTML. We could extend these controllers to accept requests from outside that are coming from OJS side. To send requests, we used jQuery RESTful calls in the JavaScript parts of the implementation.

On the OJS side we used the plugin API to implement a "general plugin". General plugins in OJS allow the inclusion of other types of plugins. This plugin can manipulate the user interface and the database, and it can be notified when a specific function or page is called in the system. We used this feature in OJS to receive a notification when an editor assigns a reviewer to a submission and to contact Fidus Writer to create, if necessary, user accounts for the reviewers, and to grant them the necessary permissions. General plugins are also capable of accepting connections from the Web. This let us develop a RESTful API for OJS to accept calls from Fidus Writer. An example of that was providing a list of journals and their identifies as the response to an HTTP GET request from Fidus Writer.

Figure 2 shows the interactions that connect the OJS and Fidus Writer of the integration system. The following clip[19] displays the interaction of Fidus Writer and OJS.

---

19  Placeholder for a link. https://github.com/OSCOSS/fiduswriter/tree/089af8b46f2de41c0224245545a47cbf217e1cf5

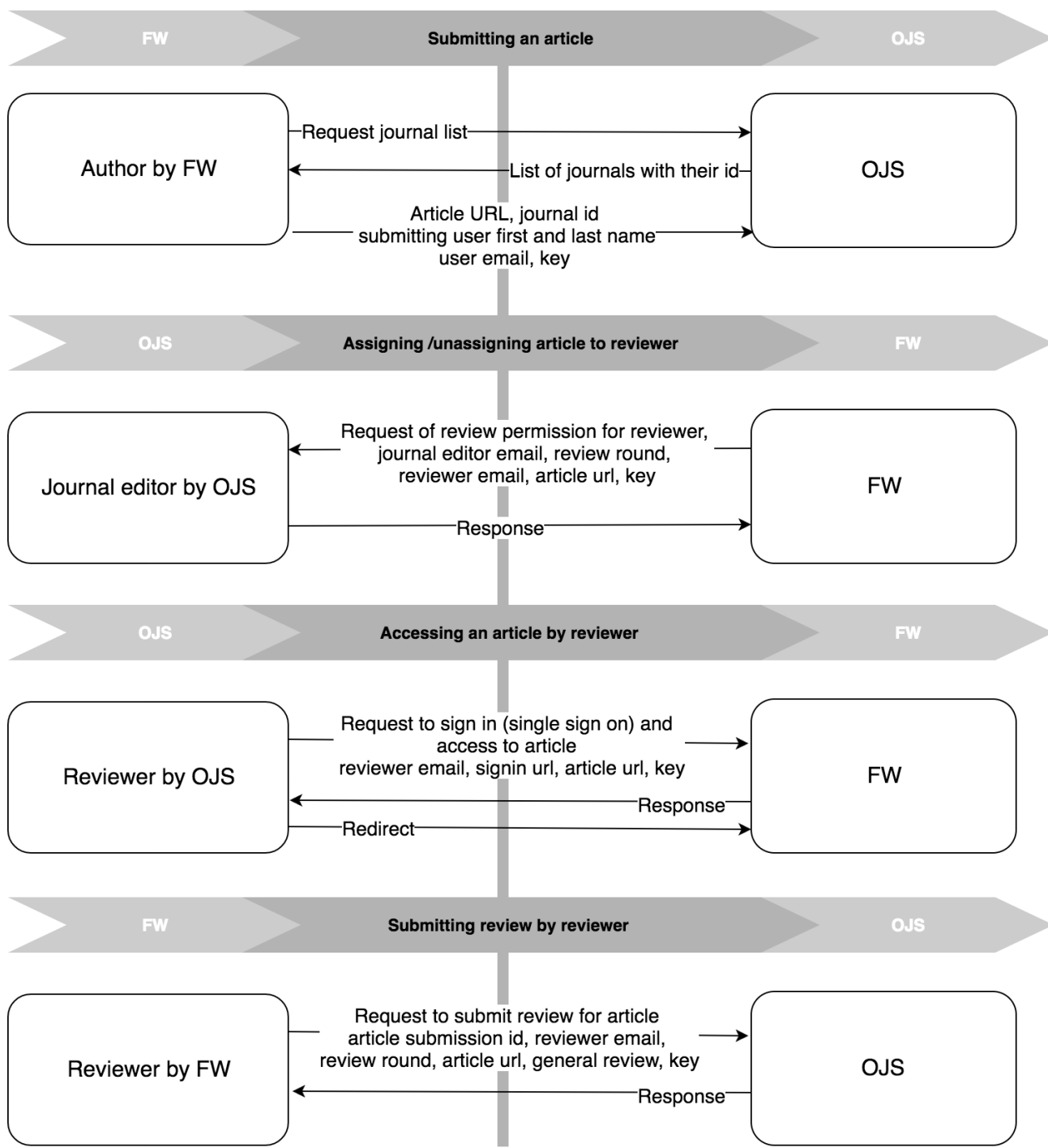

Figure 2: Interaction between Fidus Writer (FW) and OJS

# Conclusion and Outlook

We expect authors, reviewers and journal editors to be able to speed up their tasks related to authoring and reviewing manuscripts by using the integrated environment we have laid out in this article. We expect not only reduced time and thus cost but also fewer possible user mistakes (as could be made when emailing back and forth, converting, copying and uploading). Reduced administration time is another advantage of the integrated system because the accounts are created automatically and signing in one system does not make it necessary to log in the other system.

Our system is limited by the target environment that we designed our system. Over the Internet, the online connection of authors is influenced inherently by the network limitations, and the interface design is limited by being executing inside a browser. Richtext editing in browsers continues to be limited by a lack of standardization and development of relevant web technologies, as has been acknowledged even by the W3C and its member organizations (Berjon 2016).

We are planning to evaluate the system with social scientists from the communities of the mda and HSR journals who have already published articles and worked with reviewing systems. In the future, we mainly focus over following up methods to see whether the comments are affected the manuscript in order to achieve a closer integration of the authoring and the reviewing process integration.

# Acknowledgments

This work was funded by DFG, grant no. SU 647/19-1 and AU 340/9-1; the OSCOSS project at GESIS and University Bonn. We thank Fakhri Momeni for developing major parts of the Fidus Writer integration with OJS. We would like to thank Aleksandr Korovin, who implemented part of the comment visibility permissions in Fidus Writer, and Firas Kassawat for helping with integrating tables. We also thank Alec Smecher, the main developer of OJS, for helpful discussions and answering our questions regarding OJS integration API.

# Author information

Afshin Sadeghi[20] sadeghi@cs.uni-bonn.de
Affiliation: University of Bonn, Römerstraße 164, 53117 Bonn

Johannes Wilm, Philipp Mayr
Affiliation: GESIS - Leibniz Institute for the Social Sciences, Unter Sachsenhausen 6-8, 50667 Köln
philipp.mayr@gesis.org, mail@johanneswilm.org

Christoph Lange
Affiliation: University of Bonn & Fraunhofer IAIS Schloss Birlinghoven, 53757 Sankt Augustin

---

[20] Corresponding author

# References

1. Berjon, R. ed., 2016. HTML Editing Task Force Charter. Available at: http://w3c.github.io/editing/tf-charter.html.

2. Bornmann, L., 2011. Scientific peer review. Annual Review of Information Science and Technology, 45, pp.197--245. Available at: http://doi.wiley.com/10.1002/aris.2011.1440450112.

3. Bornmann, L. & Daniel, H.-D., 2010. The manuscript reviewing process: Empirical research on review requests, review sequences, and decision rules in peer review. Library & Information Science Research, 32, pp.5--12. Available at: http://linkinghub.elsevier.com/retrieve/pii/S0740818809001418.

4. Mayr, P. & Lange, C., 2016. The Opening Scholarly Communication in Social Sciences project OSCOSS. Available at: http://arxiv.org/abs/1611.04760.

5. Murakami, S. & Wilm, J., 2015. Web browser based CSS typesetting engine - How browser based typesetting systems can be made to work also for printed media. In C. Foster, ed. XML London 2015 proceedings.

6. Perkel, J.M., 2014. Scientific writing: the online cooperative. Nature, 514, pp.127--128. Available at: http://www.nature.com/doifinder/10.1038/514127a.

7. Smecher, A., 2008. The future of the electronic journal. NeuroQuantology, 6, pp.1--6.

8. Sompel, H. Van de et al., 2004. Rethinking Scholarly Communication. D-Lib Magazine, 10. Available at: http://www.dlib.org/dlib/september04/vandesompel/09vandesompel.html.

9. Whitehead, E.J., 2005. Collaborative Authoring on the Web: Introducing WebDAV. Bulletin of the American Society for Information Science and Technology, 25, pp.25--29. Available at: http://doi.wiley.com/10.1002/bult.107.

10. Wilm, J., 2015. Does Digital Publishing Need Standards? A History of Text File Standards and Outlook into the Future S. Worthington, ed. Dossier: Standards in Digital Publishing - Practitioners' Viewpoints. Available at: https://research.consortium.io/docs/publishing_standards_dossier/publishing_standards_dossier.html.

11. Wilm, J. & Frebel, D., 2015. Real-world Challenges to Collaborative Text Creation. In Proceedings of the 2Nd International Workshop on (Document) Changes: Modeling, Detection, Storage and Visualization. DChanges '14. Fort Collins, CO, USA: ACM, p. 8:1--8:4. Available at: http://doi.acm.org/10.1145/2723147.2723154.